\documentclass[twoside]{article}

\newcommand{\bs}{\boldsymbol}
\newcommand{\bfe}{\bs{e}}
\newcommand{\bfw}{\bs{\omega}}
\newcommand{\bfth}{\bs{\theta}}
\newcommand{\der}{\text{d}}
\newcommand{\defeq}{\equiv}

\newcommand{\ie}{\textit{i.e.}}
\newcommand{\eg}{\textit{e.g.}}

\usepackage{artalt}
\usepackage{amsmath}
\usepackage{latexsym}

\begin{document}

\title{Axistationary perfect fluids -- a tetrad approach}
\author{\textsc{Gyula Fodor$^\S$\thanks{E-mail:
      gfodor@rmki.kfki.hu}, Mattias Marklund$^\P$\thanks{E-mail:
      mattias.marklund@physics.umu.se}, and
    Zolt{\'a}n Perj{\'e}s$^\S$\thanks{E-mail:
      perjes@rmki.kfki.hu}}\\[3mm]
\textit{$^\S$KFKI Research Institute for Particle and Nuclear
  Physics,} \\
  \textit{Budapest 114, P.O.\ Box 49, H-1525 Hungary}\\[2mm]
  \textit{$^\P$Department of Plasma Physics, Ume{\aa} University, S-901 87
  Ume{\aa}, Sweden}}
\maketitle
\begin{abstract}
  Stationary axisymmetric perfect fluid space-times are investigated
  using the curvature description of geometries. Attention is focused
  on space-times with a vanishing electric part of the Weyl tensor.
  It is shown that the only incompressible axistationary magnetic perfect
  fluid is the interior Schwarzschild solution. The existence of a
  rigidly rotating perfect fluid, generalizing the interior Schwarzschild
  metric is proven. Theorems are stated on Petrov types and
  electric/magnetic Weyl tensors.
\end{abstract}

\section{Introduction}
   A well-known property of the static, incompressible interior
Schwarzschild space-time is conformal flatness, {\ie},
the Weyl tensor vanishing: $C_{ijk\ell}=0$.
The Weyl tensor can be split into two parts, the \emph{electric} and
\emph{magnetic} parts. The first part represents the tidal forces of
the curvature, while the second part (which has no Newtonian
analogue) is generated by the vorticity, shear, etc.\ of the
congruence to which the electric and magnetic parts are related.
Therefore, rotating matter can be viewed as partly responsible for the
existence of the magnetic part of the curvature \cite{DeFelice-Clarke}.
A simple example of a space-time containing rotating
matter is expected to be a perfect fluid with a purely magnetic Weyl
tensor.

  Little is known about space-times with a magnetic curvature. Arianrhod
{\em et al.} \cite{Arianrhod} have obtained an example of these space-times,
albeit with an unknown equation of state. Another solution, suggested in
\cite{Misra} is, however, shown to be not purely magnetic by McIntosh
\cite{McIntosh}.

To get some more insight on the subject, in this paper we
investigate perfect fluid space-times with axistationary symmetry.
We confine our attention to circularly rotating matter, {\em i.e.}, the
four-velocity lying in the plane of the two Killing vectors. Thus
we may choose the line element in its Papapetrou form,
\begin{equation}\label{eq:metric}
  {\der}s^2 = f\left({\der}t + \omega\,{\der}\varphi\right)^2
            - f^{-1}\left[\text{e}^{2\lambda}\left(
            {\der}r^2 + {\der}z^2\right)
            + \varrho^2\,{\der}\varphi^2\right] \ .
\end{equation}
We choose the canonical frame
\begin{equation}\label{eq:frame}
  \begin{array}{rll}
    & \bfth^0 \defeq f^{1/2}\left({\der}t + \omega\,{\der}\varphi\right)
     & \bfth^1 \defeq f^{-1/2}\text{e}^{\lambda}{\der}r \\
    & \bfth^2 \defeq f^{-1/2}\text{e}^{\lambda}{\der}z &
      \bfth^3 \defeq f^{-1/2}\varrho\,{\der}\varphi \ .
  \end{array}
\end{equation}
and project all quantities onto this.

A space-time is locally completely determined by the set
\begin{equation}
  S \defeq \left\{R_{ijk\ell}, \gamma^a\!_{bk},
           {\bfe}_ix^{\alpha}, \eta_{ij}\right\} \ ,
\end{equation}
where $R_{ijk\ell}$ are the components of the Riemann tensor,
$\gamma^a\!_{bk}$ are the rotation coefficients,
${\bfe}_ix^{\alpha}$ the gradients of the essential coordinates,
and $\eta_{ij}$ is the constant metric in the chosen
frame\footnote{Note that the analysis would be possible in any
  constant frame, not just the Lorentz frame. We choose this
  particular frame since it is related, in a simple fashion, to the
  physical observables of the fluid.}
\cite{Bradley-Karlhede,Bradley-Marklund}.
The set $S$ may be calculated using the frame (\ref{eq:frame})
and has a form characteristic of the metric
(\ref{eq:metric}). We still have some freedom in performing Lorentz
transformations of the frame (\ref{eq:frame}), {\eg}, rotations in the
(1,2)-plane (see Sec.\ \ref{sec:equations}). These transformations
will change the appearance of the components of, say, the Riemann
tensor, but will not introduce new {\em non-zero} components of this
tensor. This freedom can be exploited to make the set $S$ as simple as
possible. There is also a boost freedom, which does not introduce any
new non-zero components of the tensors in the set $S$. This can
be used to put the frame (\ref{eq:frame}) in a comoving form, thus
giving a natural interpretation of (some of) the Ricci rotation
coefficients in terms of fluid variables (such as vorticity).

If we start with a set $S$, instead of generating the set $S$ from a
metric satisfying Einstein's field equations, we need to impose
certain restrictions on the elements of the set in order to guarantee
that $S$ can be generated by a line-element
that satisfy Cartan's equations under the restriction of
Einstein's field equations. In general, these conditions are part of
the commutator equations, Ricci equations, and Bianchi equations
\cite{Bradley-Karlhede,Bradley-Marklund}.

The approach taken in this paper proceeds in the following
steps:
\begin{enumerate}
  \item Investigate the form of $S$ in the Lorentz frame
    (\ref{eq:frame}) for the metric (\ref{eq:metric}).
  \item Analyze the properties of this $S$ under Lorentz
    transformations.
  \item Using the knowledge obtained in the steps above, make an
    \emph{ansatz} for the set $S$.
  \item Subject the elements of this Ansatz to the necessary and
    sufficient equations (in general, parts of the commutator
    equations, Ricci equations, and Bianchi equations).
  \item Use the fact that part of the variables in $S$ are the
    components of the Weyl tensor to investigate magnetic
    space-times.
\end{enumerate}

In Sec.\ \ref{sec:equations} we derive the equations describing an
axisymmetric and stationary perfect fluid space-time. These are the
commutators acting on the ``essential coordinates'' (on which all
quantities in $S$ depend) and the Ricci equations. The Bianchi
equations are not needed, but in practice they are very useful when
there are restrictions on the Weyl tensor.
All these equations are first order partial differential equations in
terms of two variables for the quantities in the set $S$. Also, $S$
is chosen as to be compatible with the metric (\ref{eq:metric}) in a
Lorentz transformed version of the frame (\ref{eq:frame}).

In Sec.\ \ref{sec:magnetic} we investigate the restrictions of 
Sec.\ \ref{sec:equations} in the case of
a purely magnetic Weyl tensor. Specifically, we
examine what space-times are allowed under the further restriction of
incompressible fluid (constant energy density) or shear-free motion
of the fluid (rigid rotation).

Section \ref{sec:metric} describes how to
obtain the full metric of the space-time. We focus on the
physically important shear free space-times. We show that the 
line-element can be written in terms of quadratures once the
elements of the set $S$ are determined by the equations set up in
Sec.\ \ref{sec:equations}.

In Sec. \ref{sec:petrov}, we consider various Petrov types of circularly
rotating perfect fluids. We prove that type N fluids have the unphysical
equation of state $\mu+p=0$, and that rigidly rotating purely electric
fluids are of Petrov type D.

\section{Equations}\label{sec:equations}

Our space-time possesses two Killing vectors, one time-like and one
space-like, which commute. We assume that the coordinates are adapted
to the Killing vectors, such that all quantities in the set $S$ are
dependent on two coordinates $\{x^{\alpha}\} \defeq \{x, y\}$.

The source is a perfect fluid with the energy-momentum tensor
\begin{equation}
  T_{ij} = (\mu + p)u_iu_j - pg_{ij} \ .
\end{equation}

We choose a frame ${\bfe}_i \defeq
(\bfe_ix^{\mu})({\partial}/{\partial}x^{\mu})$ with dual
${\bfw}^i \defeq \omega^i\!_{\mu}\,{\der}x^{\mu}$, $i = 0, ..., 3$,
such that
\begin{align}
  & \der s^2 = ({\bfw}^0)^2 - ({\bfw}^1)^2 - ({\bfw}^2)^2
               - ({\bfw}^3)^2   \label{eq:frame1} \\
  & {\bfe}_0 \defeq {\bs{u}} \label{eq:frame2} \ ,
\end{align}
{\ie}, the frame is comoving and $(g_{ij}) = (\eta_{ij}) \defeq
\text{diag}(1, -1, -1, -1)$.
In terms of the frame (\ref{eq:frame}) related to the metric
(\ref{eq:metric}), the choice of a comoving frame means that we have
carried out a boost in the (0,3)-plane:
\begin{equation}\label{boost}
  \begin{pmatrix}
   {\bfth}^0 \\
   {\bfth}^1 \\
   {\bfth}^2 \\
   {\bfth}^3
  \end{pmatrix} \mapsto
  \begin{pmatrix}
    \cosh\chi & 0 & 0 & \sinh\chi \\
    0           & 1 & 0 & 0 \\
    0           & 0 & 1 & 0 \\
    \sinh\chi & 0 & 0 & \cosh\chi
  \end{pmatrix}
  \begin{pmatrix}
    {\bfth}^0 \\
    {\bfth}^1 \\
    {\bfth}^2 \\
    {\bfth}^3
  \end{pmatrix}  \ ,
\end{equation}
or explicitly
\begin{align}
  & {\bfth}^0 \mapsto \bfth^0 = {\cosh\chi}f^{1/2}({\der}t +
  \omega\,\der\varphi) + {\sinh\chi}f^{-1/2}\varrho\,\der\varphi \\
  & {\bfth}^3 \mapsto \bfth^3 =
  {\cosh\chi}f^{-1/2}\varrho\,\der\varphi + {\sinh\chi}f^{1/2}({\der}t
  + \omega\,\der\varphi) \ ,
\end{align}
where $\chi = \chi(x,y)$.
In the case of a rigidly rotating fluid, the boost parameter $\chi$
may always be put to zero. With this, the contravariant vectors
$\bfe_0$ and $\bfe_3$ become
\begin{align}
  & \bfe_0 = \frac{1}{\varrho}\left[\left(\sinh\chi f^{1/2}\omega +
  \cosh\chi f^{-1/2}\varrho\right)\partial_t -
  \sinh\chi f^{1/2}\partial_{\varphi} \right] \label{e0} \\
  & \bfe_3 = \frac{1}{\varrho}\left[-\left(\cosh\chi f^{1/2}\omega +
  \sinh\chi f^{-1/2}\varrho\right)\partial_t +
  \cosh\chi f^{1/2}\partial_{\varphi} \right] \ . \label{e3}
\end{align}

We define the non-zero components of the Riemann tensor of the circularly
rotating fluid in the frame
given by Eqs.\ (\ref{eq:frame1}) and (\ref{eq:frame2}) as
\begin{alignat*}{2}
  & R_{0101} = E_1 - \tfrac16(\mu + 3p)   & \quad &
    R_{0102} = E_3   \\
  & R_{0113} = -H_3   & \quad & R_{0123} = H_1   \\
  & R_{0202} = E_2 - \tfrac16(\mu + 3p)   & \quad &
    R_{0213} = -H_2   \\
  & R_{0223} = H_3   & \quad
  & R_{0303} = -E_1 - E_2 - \tfrac16(\mu + 3p)   \\
  & R_{0312} = -H_1 - H_2   & \quad
  & R_{1212} = E_1 + E_2 - \tfrac13\mu   \\
  & R_{1313} = -E_2 - \tfrac13\mu   & \quad
  & R_{1323} = E_3   \\
  & R_{2323} = -E_1 - \tfrac13\mu \ , & \quad &
\end{alignat*}
where $E_{A}$ and $H_A$, $A = 1, 2, 3$, are components of
the electric part $E_{ij} \defeq C_{ikj\ell}u^ku^{\ell}$ and magnetic
part $H_{ij} \defeq {}^*\!C_{ikj\ell}u^ku^{\ell}$ of the Weyl tensor
respectively. The rotation coefficients $\gamma^i\!_{jk} \defeq
\omega^i\!_{\mu;\nu}(\bfe_jx^{\mu})(\bfe_kx^{\nu})$ are
\begin{alignat*}{2}
  & \gamma_{010} = -a_1   & \quad & \gamma_{020} = -a_2   \\
  & \gamma_{013} = -\sigma_2 - \omega_2   & \quad &
    \gamma_{023} = \sigma_1 + \omega_1   \\
  & \gamma_{031} = -\sigma_2 + \omega_2   & \quad &
    \gamma_{032} = \sigma_1 - \omega_1   \\
  & \gamma_{121} \defeq \alpha_1   & \quad &
    \gamma_{122} \defeq -\alpha_2   \\
  & \gamma_{130} \defeq \gamma_2   & \quad &
    \gamma_{133} \defeq \beta_1   \\
  & \gamma_{230} \defeq -\gamma_1  & \quad &
    \gamma_{233} \defeq \beta_2
\end{alignat*}
where $a_i$ is the acceleration, $\sigma_i$ the shear, and $\omega_i$
the
vorticity of the fluid.

We still have the freedom to perform rotations
\begin{equation}\label{transf}
  \begin{pmatrix}
   {\bfw}^0 \\
   {\bfw}^1 \\
   {\bfw}^2 \\
   {\bfw}^3
  \end{pmatrix} \mapsto
  \begin{pmatrix}
    1 & 0 & 0 & 0 \\
    0 & \cos\vartheta  & \sin\vartheta & 0 \\
    0 & -\sin\vartheta & \cos\vartheta & 0 \\
    0 & 0 & 0 & 1
  \end{pmatrix}
  \begin{pmatrix}
    {\bfw}^0 \\
    {\bfw}^1 \\
    {\bfw}^2 \\
    {\bfw}^3
  \end{pmatrix}
\end{equation}
in the (1,2)-plane (here $\vartheta = \vartheta(x,y)$).\footnote{
During this transformation, the rotation coefficients transform as
\[
  \hat{\gamma}^i\!_{jk} = \Lambda^i\!_{\ell}\Lambda_j\!^m\Lambda_k\!^n%
  \gamma^{\ell}\!_{mn}
  + (\bfe_n\Lambda^i\!_{\ell})\Lambda_j\!^{\ell}\Lambda_k\!^n \ ,
\]
where $\Lambda^i\!_{\ell}$ is the transformation matrix and
$\Lambda^i\!_{\ell}\Lambda_j\!^{\ell} = \delta^i_j$.} This rotation
will not introduce any new non-zero components in the Riemann tensor
or the rotation coefficients (as any other rotation would do), although
the components of the respective geometric objects will
mix.

Parts of the Ricci equations and Bianchi equations are solved by the
choice
${\bfe}_0x = {\bfe}_0y = {\bfe}_3x = {\bfe}_3y = 0$. Using the reduced
number of coordinate gradients, we may write
\begin{align}
  & \bfe_1 = \xi_1\partial_x + \upsilon_1\partial_y  \\
  & \bfe_2 = \xi_2\partial_x + \upsilon_2\partial_y \ ,
\end{align}
where $\xi_i \defeq \bfe_ix$, $\upsilon_i \defeq \bfe_iy$.

The commutator equations are generated by
\begin{equation}\label{comgen}
  [\bfe_i,\bfe_j]=(\gamma^k\!_{ji} -\gamma^{k}\!_{ij})\bfe_k \ .
\end{equation}
We will refer to a specific commutator by the index pair $(i,j)$.
Acting with the $(1,2)$ commutator on $x$ and $y$,
we obtain the \\[2mm]
\textit{commutator equations}
\begin{align}\label{comx}
  & \bfe_1\xi_2 - \bfe_2\xi_1 =
     -\alpha_1\xi_1 +\alpha_2\xi_2  \\
  & \bfe_1\upsilon_2 - \bfe_2\upsilon_1 =
     -\alpha_1\upsilon_1
     +\alpha_2\upsilon_2\label{comy} \ .
\end{align}
Writing $\bfe_0$ and $\bfe_3$ as linear combinations of $\partial_t$
and $\partial_\varphi$, from the $(0,3)$ commutators acting on $x$
and $y$ follows that
\begin{align}
  & \gamma_1 = \sigma_1 + \omega_1   \\
  & \gamma_2 = \sigma_2 + \omega_2 \ ,
\end{align}
respectively. All the remaining commutator equations are identically
satisfied when acting on the essential coordinates $x$ and $y$.

From  the relation $R^a\!_{bij} =
{\bfe}_i\gamma^a\!_{bj} - {\bfe}_j\gamma^a\!_{bi}+
2\gamma_{kb[i}\gamma^{ak}\!_{j]} + 2\gamma^a\!_{bk}\gamma^k\!_{[ij]}$
we obtain the\\[2mm]
\textit{Ricci equations}
\begin{align}
 \bfe_1a_1 &=E_1-\tfrac 16(\mu +3p)+a_1^2+a_2\alpha_1
           -(\sigma_2+\omega_2)(3\sigma_2-\omega_2)
\label{rr1}\\
 \bfe_2a_2 &=E_2-\tfrac 16(\mu +3p)+a_2^2+a_1\alpha_2
           -(\sigma_1+\omega_1)(3\sigma_1-\omega_1)
\label{rr2}\\
 \bfe_2a_1 &=E_3+a_1a_2-a_2\alpha_2
           +(3\sigma_1-\omega_1)(\sigma_2+\omega_2)
\label{rr3}\\
 \bfe_1a_2 &=E_3+a_2a_1-a_1\alpha_1
           +(3\sigma_2-\omega_2)(\sigma_1+\omega_1)
\label{rr4}\\
 \bfe_1(\sigma_1+\omega_1) &=-H_2-2a_2\omega_2
   +\alpha_1(\sigma_2+\omega_2)-\beta_1(\sigma_1-\omega_1)
   +\beta_2(\sigma_2-\omega_2)
\label{rr5}\\
 \bfe_2(\sigma_2+\omega_2) &=-H_1-2a_1\omega_1
   +\alpha_2(\sigma_1+\omega_1)-\beta_2(\sigma_2-\omega_2)
   +\beta_1(\sigma_1-\omega_1)
\label{rr6}\\
 \bfe_2(\sigma_1+\omega_1) &=H_3+2a_2\omega_1
   -\alpha_2(\sigma_2+\omega_2)-2\beta_2\sigma_1
\label{rr7}\\
 \bfe_1(\sigma_2+\omega_2) &=H_3+2a_1\omega_2
   -\alpha_1(\sigma_1+\omega_1)-2\beta_1\sigma_2
\label{rr8}\\
 \bfe_1\beta_1 &=-E_2-\tfrac 13\mu +\alpha_1\beta_2-\beta_1^2
   -(\sigma_2-3\omega_2)(\sigma_2+\omega_2)
\label{rr9}\\
 \bfe_2\beta_2 &=-E_1-\tfrac 13\mu +\alpha_2\beta_1-\beta_2^2
   -(\sigma_1-3\omega_1)(\sigma_1+\omega_1)
\label{rr10}\\
 \bfe_2\beta_1 &=E_3-\alpha_2\beta_2-\beta_1\beta_2
   +(\sigma_1-3\omega_1)(\sigma_2+\omega_2)
\label{rr11}\\
 \bfe_1\beta_2 &=E_3-\alpha_1\beta_1-\beta_2\beta_1
   +(\sigma_2-3\omega_2)(\sigma_1+\omega_1)
\label{rr12}
\end{align}
and
\begin{align}
 \bfe_1(\sigma_1-\omega_1)& + \bfe_2(\sigma_2-\omega_2) =
   -H_1 - H_2 + \alpha_1(\sigma_2 - \omega_2)
   + \alpha_2(\sigma_1 - \omega_1)
\label{rr13}\\
 \bfe_2\alpha_1 + \bfe_1\alpha_2 &= -E_1 - E_2 + \tfrac 13\mu
   + \alpha_1^2 + \alpha_2^2
\label{rr14}\\
  E_1 + E_2 &= -\tfrac 16(\mu + 3p) - a_1\beta_1 - a_2\beta_2
   + (\sigma_1 + \omega_1)^2 + (\sigma_2 + \omega_2)^2
\label{rr15}\\
  H_1 + H_2 &= -(a_1 + \beta_1)(\sigma_1 + \omega_1)
   - (a_2 + \beta_2)(\sigma_2 + \omega_2)  \ .
\label{rr16}
\end{align}

The Bianchi equations $R_{ij[k\ell;m]} = 0$ are not necessary
equations (since the space-time is without isotropy
\cite{Bradley-Karlhede, Bradley-Marklund}), but they may be used as
``help'' equations (although all information in them are contained
in the lower-derivative stage of the Ricci equations for
stationary axisymmetric space-times). Thus we have the \\[2mm]
\textit{twice contracted Bianchi equations $T^{ij}\!_{;j} = 0$}
\begin{align}
  &\bfe_1p = a_1(\mu + p) \label{bi1}  \\
  &\bfe_2p = a_2(\mu + p) \label{bi2} \ ,
\end{align}
and the \\[2mm]
\textit{Bianchi equations}
\begin{align}
 \bfe_1(E_1+E_2) &= (a_1-2\beta_1)E_1+(2a_1-\beta_1)E_2
   -(a_2+\beta_2)E_3 \nonumber\\
 &+(\sigma_1+\omega_1)(H_1-H_2)
   +2(\sigma_2+\omega_2)H_3-\tfrac 16e_1\mu
\label{b1}\\
 \bfe_2(E_2+E_1) &= (a_2-2\beta_2)E_2+(2a_2-\beta_2)E_1
   -(a_1+\beta_1)E_3 \nonumber\\
 &+(\sigma_2+\omega_2)(H_2-H_1)
   +2(\sigma_1+\omega_1)H_3-\tfrac 16e_2\mu
\label{b2}\\
 \bfe_1E_1+e_2E_3 &= (\alpha_2-2\beta_1)E_1
   -(\alpha_2+\beta_1)E_2+(2\alpha_1-\beta_2)E_3 \nonumber\\
 &+(\sigma_1+3\omega_1)H_1+2\sigma_1H_2
   -(\sigma_2-3\omega_2)H_3-\tfrac 13e_1\mu
\label{b3}\\
 \bfe_2E_2+e_1E_3 &= (\alpha_1-2\beta_2)E_2
   -(\alpha_1+\beta_2)E_1+(2\alpha_2-\beta_1)E_3 \nonumber\\
 &+(\sigma_2+3\omega_2)H_2+2\sigma_2H_1
   -(\sigma_1-3\omega_1)H_3-\tfrac 13e_2\mu
\label{b4}
\end{align}
and
\begin{align}
 \bfe_1(H_1+H_2) &= (a_1-2\beta_1)H_1
   +(2a_1-\beta_1)H_2-(a_2+\beta_2)H_3 \nonumber\\
 &+(\sigma_1+\omega_1)(E_2-E_1)-2(\sigma_2+\omega_2)E_3
   +\tfrac 12(\sigma_1+\omega_1)(\mu +p)
\label{b5}\\
 \bfe_2(H_2+H_1) &= (a_2-2\beta_2)H_2
   +(2a_2-\beta_2)H_1-(a_1+\beta_1)H_3 \nonumber\\
 &+(\sigma_2+\omega_2)(E_1-E_2)-2(\sigma_1+\omega_1)E_3
   +\tfrac 12(\sigma_2+\omega_2)(\mu +p)
\label{b6}\\
 \bfe_1H_1+e_2H_3 &= (\alpha_2-2\beta_1)H_1
   -(\alpha_2+\beta_1)H_2+(2\alpha_1-\beta_2)H_3 \nonumber\\
 &-(\sigma_1+3\omega_1)E_1-2\sigma_1E_2
   +(\sigma_2-3\omega_2)E_3+\omega_1(\mu +p)
\label{b7}\\
 \bfe_2H_2+e_1H_3 &= (\alpha_1-2\beta_2)H_2
   -(\alpha_1+\beta_2)H_1+(2\alpha_2-\beta_1)H_3 \nonumber\\
 &-(\sigma_2+3\omega_2)E_2-2\sigma_2E_1
   +(\sigma_1-3\omega_1)E_3+\omega_2(\mu +p) \ .
\label{b8}
\end{align}

Once more, we stress the fact that in order to find a solution to
Einstein's equations, we need to solve Eqs.\ (\ref{comx}),
(\ref{comy}), and (\ref{rr1})--(\ref{rr16}). The use of the
 Bianchi equations (\ref{b1})--(\ref{b8}) is optional when no 
further symmetry restrictions are imposed.

\section{Magnetic space-times}\label{sec:magnetic}
 For space-times with a purely magnetic Weyl tensor, we set $E_A=0$.

\subsection{Incompressible fluids}
Incompressible fluids are defined by $\bfe_1\mu = \bfe_2\mu = 0$.
We use the rotation (\ref{transf}) to set $H_3 = 0$. We may then
prove the following:\\[2mm]
{\bf Theorem 1.}
{\it The only incompressible axistationary perfect fluid with a
pure magnetic  Weyl tensor is the interior Schwarzschild
solution.}\\[2mm]
{\it Proof.}
  From Eqs.\ (\ref{b1})--(\ref{b4}) we get four algebraic equations
from which $H_1=H_2$ and $\sigma_1+\omega_1=\sigma_2+\omega_2=0$.
But Eq.\ (\ref{rr16}) then results in $H_1 = H_2 = 0$, {\ie},
the space-time is conformally flat. According to a theorem by Collinson
\cite{Collinson}, the only conformally flat axistationary perfect fluid
space-time is the interior Schwarzschild solution. \hfill
$\Box$\\[2mm]

\subsection{Rigidly rotating fluids}
When the fluid is rigid, we have $\sigma_1 = \sigma_2 = 0$. We set
$\omega_1
= 0$ by using the transformation (\ref{transf}). Then the Bianchi
equations
(\ref{b1})--(\ref{b4}) give the constraints
\begin{equation}
  \omega_2H_3 = 0 \ , \quad \omega_2(2H_1 + H_2) = 0 \ .
\end{equation}

If $\omega_2 = 0$, the fluid is static and Eqs.\ (\ref{b1}) and
(\ref{b2})
imply $\bfe_1\mu = \bfe_2\mu = 0$, while Eqs.\ (\ref{rr5}),
(\ref{rr7}), and
(\ref{rr6}) gives  $H_1 = H_2 = H_3 = 0$. Thus, this is the interior
Schwarzschild solution.

If $H_3 = 0$ and $H_2 = -2H_1$, we obtain a consistent subsystem of
equations
\begin{align}
  & \bfe_1p =\bfe_1\mu = \bfe_1\beta_2 = \bfe_1a_2 = 0 \label{a21}\\
  & \bfe_2a_2 = a_2^2 - \tfrac16(\mu + 3p) \label{a22} \\
  & \bfe_2\beta_2 = -\tfrac13\mu - \beta_2^2 \label{phi2} \\
  & \bfe_2\mu = -3(a_2 + \beta_2)(\mu + 3p + 6a_2\beta_2)\label{mu2}\\
  & \bfe_2p = a_2(\mu + p) \label{p2} \\
  & \bfe_1\beta_1 = -b^2- \beta_1^2 \label{delta1}\\
  & \bfe_2\beta_1 = -\beta_1\beta_2 \ , \label{delta2}
\end{align}
where $b^2 \defeq \beta_2^2 - 3a_2\beta_2 - \tfrac16\mu - \tfrac32p$.
(For the interior Schwarzschild solution $b^2>0$.)
This system is characterized by $\omega_1 = a_1 = \alpha_2 = 0$,
$\alpha_1 = -\beta_2$, $H_1 = (a_2 + \beta_2)\omega_2$, and
$\omega_2^2 =(\mu + 3p)/6 + a_2\beta_2$. We choose coordinates
$\{x, y\}$ such that
$\bfe_1 =\zeta(\partial/{\partial}x)$, where we assume that
$\zeta\equiv\zeta(y)$ depends only on $y$, and
$\bfe_2 = \xi(\partial/{\partial}x) + \upsilon(\partial/{\partial}y)$.
With this choice of coordinates, all quantities except $\beta_1$
become independent of $x$.
We still have the freedom to make the coordinate transformations
\begin{align}
  & \bar{x} = h_1(y) x + h_0(y) \label{coord1} \\
  & \bar{y} = \bar{y}(y) \label{coord2} \ ,
\end{align}
where $h_1(y)$ and $h_0(y)$ are some $x$-independent functions.
It follows from the $(1,2)$ commutator on $y$ ({\em cf.}\ Eq.\
(\ref{comy})), that $\upsilon$ depends only on $y$. Acting with the
commutator $(1,2)$ on $x$ (\emph{cf.}\ Eq.\ (\ref{comx})), we find that
\begin{equation}
 \zeta\frac{\partial\xi}{\partial x}
 -\upsilon\frac{\partial\zeta}{\partial y}=\zeta\beta_2 \ .
\end{equation}
Integrating, we obtain
\begin{equation}\label{eqxi}
  \xi=\left(\beta_2+\frac v\zeta\frac{\partial\zeta}{\partial y}
  \right)x +\xi_0 \ ,
\end{equation}
where $\xi_0$ is a function depending only on the coordinate $y$.
Performing a transformation (\ref{coord1}), $\xi$ becomes
\begin{equation}\label{xitran}
\bar\xi=\xi\,h_1(y)+\upsilon\left(x\frac{dh_1(y)}{dy}
+\frac{dh_0(y)}{dy}\right) \ .
\end{equation}
Substituting (\ref{eqxi}), we can achieve $\bar\xi=0$ by suitably
choosing $h_1(y)$ and $h_0(y)$. This means that we can use a
coordinate system in which
\begin{equation}
  \bfe_1=\zeta\frac{\partial}{\partial x} \ ,\qquad
  \bfe_2=\upsilon\frac{\partial}{\partial y} \ ,
\end{equation}
and both $\zeta$ and $\upsilon$ depend only on $y$.
Since $b$ does not depend on $x$ either, we can integrate
(\ref{delta1}) to obtain
\begin{equation}
  \beta_1 = -b\tan\left[\frac b\zeta(x-x_0)\right] \ ,
\end{equation}
where $x_0$ is some function of $y$.
Substituting $\xi$ and $\beta_1$ in Eq.\ (\ref{delta2}) and using
$\bfe_2 b=-\beta_2 b$ we get that $x_0$ must be a constant. Since
we only used the derivative of $h_0(y)$ in the transformation 
(\ref{xitran}), we still may add a constant to the coordinate $x$,
and we use this freedom to set $x_0=0$.

The remaining equations are
\begin{align}
  & \upsilon\frac{da_2}{dy} = a_2^2 - \tfrac16(\mu + 3p) \label{ups1} \\
  & \upsilon\frac{d\beta_2}{dy} = -\tfrac13\mu - \beta_2^2 \\
  & \upsilon\frac{d\mu}{dy} = -3(a_2 + \beta_2)
    (\mu + 3p + 6a_2\beta_2) \\
  & \upsilon\frac{dp}{dy} = a_2(\mu + p)\label{ups4} \ .
\end{align}
By introducing $B \defeq b^2$, $\Omega \defeq \omega_2^2$,
these equations can also be written as
\begin{align}
  & \upsilon\frac{dB}{dy} = -2B\beta_2 \label{Bder}\\
  & \upsilon\frac{d\beta_2}{dy} = B - 3\Omega \\
  & \upsilon\frac{da_2}{dy} = a_2(a_2 + \beta_2) - \Omega \\
  & \upsilon\frac{d\Omega}{dy} = -2\Omega(a_2 + 2\beta_2)
\end{align}

The equation of state $\mu=\mu(p)$ is given by a lengthy third-order
nonlinear differential equation. $\mu=\text{constant}$ is the interior
Schwarzschild solution (see Theorem 1).

We see that in the static limit $\omega_2 \rightarrow 0$, we
obtain $H_1 = H_2 = 0$ and $\bfe_1\mu = \bfe_2\mu = 0$. Thus in
this limit we obtain the interior Schwarzschild solution.
The acceleration $a_2$ is expressible in terms of $p$:
\begin{equation}
  a_2^2 = \tfrac23\mu + p + D(\mu + p)^2 \ ,
\end{equation}
where $D$ is a constant. For spherical symmetry, the quantities in the
set $S$ are invariant under the action of a 3-dimensional isotropy
group at the center of the spheres. Thus, denoting center values by
an index $c$, $(a_2)_c = 0$,\footnote{This follows since $a_2$ is
  the 2-component of the acceleration covector which must be zero at
  the center, otherwise there is a preferred direction.} and
\begin{equation}
  D = -\frac13\frac{2\mu_c + 3p_c}{(\mu_c + p_c)^2} \ .
\end{equation}

We collect our results in the following theorem:\\[2mm]
{\bf Theorem 2.}
{\it The only rigidly rotating axistationary perfect fluid solutions
  with a pure magnetic Weyl tensor are the interior Schwarzschild
  solution and the space-time given by Eqs.\
  (\ref{ups1})--(\ref{ups4}).}\\[2mm]

\section{Restoring the metric}\label{sec:metric}
We can always generate the set $S$ from the metric
(\ref{eq:metric}) in the boosted frame (\ref{boost}). Thus, to obtain
the full metric we only need to apply the commutators to the two
coordinates $t$ and $\varphi$ of which all geometric quantities are
 independent,
since the 1-forms $\bfw^1$ and $\bfw^2$ are known from solving the
necessary and sufficient equations. Thus, to obtain the ``missing
part'' of the metric, we must integrate
\begin{align}
  & \left[\bfe_i,\bfe_j\right]t = \left(\gamma^0\!_{ji} -
    \gamma^0\!_{ij}\right)\bfe_0t + \left(\gamma^3\!_{ji} -
    \gamma^3\!_{ij}\right)\bfe_3t \label{comm1} \\
  & \left[\bfe_i,\bfe_j\right]\varphi = \left(\gamma^0\!_{ji} -
    \gamma^0\!_{ij}\right)\bfe_0\varphi + \left(\gamma^3\!_{ji} -
    \gamma^3\!_{ij}\right)\bfe_3\varphi \ , \label{comm2}
\end{align}
where the form of the right hand side follows from the fact that
$\bfe_1$ and $\bfe_2$ are linear combinations of $\partial_x$ and
$\partial_y$. Because of the complicated form of $\bfe_i$, the
commutator equations (\ref{comm1}) and (\ref{comm2}) will in general
be difficult to integrate. This is of no physical importance, though,
since we know that the set $S$ gives a complete local description of
the space-time. The central dilemma is the condition for matching a
given fluid solution to a vacuum counterpart. This problem can in
principle be solved using the metric description, while no such
procedure exists (yet) for the description in terms of $S$.

\subsection{The shear-free case}
When the fluid is shear-free, {\ie}, rigidly rotating, we may put the
boost parameter $\chi$ equal to zero, since the frame (\ref{eq:frame})
is comoving in the rigid case. The expressions (\ref{e0}) and
(\ref{e3}) for $\bfe_0$ and $\bfe_3$, respectively, simplify
considerably:
\begin{align}
  & \bfe_0 = f^{-1/2}\partial_t \label{ee0}\\
  & \bfe_3 = \frac{f^{1/2}}{\varrho}\left(
   \partial_{\varphi}-\omega\partial_t \right)\label{ee3} \ .
\end{align}
Here $f$, $\varrho$ and $\omega$ are functions of $x$ and $y$.

We proceed to determining the $x$ (angular) dependence of the metric
quantities by use of the commutator equations (\ref{comm1}) and
(\ref{comm2}). From the $(0,1)$ commutator acting on $t$, we get
${\partial f}/{\partial x}=0$, and hence $f$ is a function of $y$
only. Integrating the equation obtained from the $(1,3)$ commutator
acting on $\varphi$, we get
\begin{equation}
  \varrho = \varrho_0\cos\bigl(\frac{b}{\zeta}x\bigr) \ ,
\end{equation}
while acting on $t$ gives
\begin{equation}
  \omega = -\frac2{bf}\omega_2\varrho_0\sin
    \bigl(\frac{b}{\zeta}x\bigr)+\omega_0 \ ,
\end{equation}
where $\varrho_0$ and $\omega_0$ depend on $y$ only.
The commutator equation $(0,2)$ acting on
$t$ and (1,2) acting on $x$ give, respectively
\begin{align}
 & a_2=-\frac\upsilon2\frac{d}{dy}\ln f  \\
 & \beta_2 = -\upsilon\frac{d}{dy}\ln\zeta \label{phi}\ .
\end{align}
Comparison with (\ref{Bder}) shows $b=C\zeta$ where one can set
the constant $C=1$ by rescaling the coordinate $x$. 
The action of $(2,3)$ on $\varphi$ yields
\begin{equation}
\zeta=C'\frac{\sqrt f}{\rho_0} \
\end{equation}
where we can arrange the constant $C'=1$ by rescaling the 
coordinate $\varphi$.

Substituting $\beta_2$ in the $(2,3)$ commutator acting on $t$ ,
the resulting equation can be separated into $x$-dependent and
$x$-independent terms. From the $x$-independent part it follows that
$\omega_0$ is a constant. Hence we can set $\omega_0=0$ by
$t\rightarrow t+\omega_0\varphi$.
The terms proportional to $\sin(bx)$ can be
integrated,
\begin{equation}
  2\omega_2 = K{\sqrt f}{\zeta^2} \
\end{equation}
where $K$ is a constant.
Putting together these results, we find $\omega=K\sin x$.
By use of the transformation (\ref{coord2}), we arrange
$\upsilon=\sqrt f$.
The space-time metric takes the form
(\emph{cf.}\ Eq.\ (\ref{eq:metric}))
\begin{equation}\label{xymetr}
 {\der}s^2 = f({\der}t + K\sin x\,\der\varphi)^2
  -\frac1f\left[{\der}y^2+\rho_0^2({\der}x^2+\cos^2 x\,\der\varphi^2
          )\right] \ .
\end{equation}

 The angular dependence of this metric is identical with the angular
dependence of the metric in Ref.\ \cite{Newman}. In that reference,
the Ansatz is used that the eigenrays of the Killing vector
$\partial/\partial t$ are geodesic. Both solutions are of NUT type.
For a full understanding of the relationship of the physical
assumptions used in both papers, a study of the radial behavior
is needed.

\section{Petrov types}\label{sec:petrov}
    In this section, we classify the circularly rotating perfect fluid
space-times into Petrov types. A Newman-Penrose null tetrad convenient for
this investigation is given by
\begin{align}
&{\bf l}=\frac{1}{\sqrt2}(\bfe_0+\bfe_3)  \\
&{\bf n}=\frac{1}{\sqrt2}(\bfe_0-\bfe_3)  \\
&{\bf m}=\frac{1}{\sqrt2}(\bfe_1+i\bfe_2) \ .
\end{align}
The nonvanishing Weyl spinor components are
\begin{align}
&\Psi_0=\frac12[E_2 - E_1 - 2H_3 + i(H_1 - H_2 - 2E_3)] \label{psi0}\\
&\Psi_2=\frac12[E_1 + E_2        - i(H_1 + H_2       )] \label{psi2}\\
&\Psi_4=\frac12[E_2 - E_1 + 2H_3 + i(H_1 - H_2 + 2E_3)] \ .
\end{align}
The algebraic Petrov type is determined by the roots of the equation
\begin{equation}
\Psi_4 b^4+6\Psi_2 b^2+\Psi_0=0\ . \label{petrov}
\end{equation}
The Petrov type is II if and only if either
$\Psi_0=0$ or $\Psi_4=0$. (Reversal of the $\bfe_3$ direction
interchanges $\Psi_0$ and $\Psi_4$.)
Petrov D type fields are characterized by the conditions
\begin{align}
&4E_1H_1 + 5E_1H_2 + 5E_2H_1 + 4E_2H_2 - 2E_3H_3 = 0 \label{typeD1}\\
&2E_1^2 + 5E_1E_2 + 2E_2^2 - E_3^2 - 2H_1^2 - 5H_1H_2 - 2H_2^2 + H_3^2 = 0
                                                    \label{typeD2} \ .
\end{align}
The type is N if $\Psi_0=\Psi_2=0$. Type III is excluded by equation
(\ref{petrov}).

From the point of view of star modeling, type N perfect fluids can hardly
be of any significance. This is due to the following \\[2mm]
{\bf Theorem 3.}
{\it A type N axistationary perfect fluid spacetime has a constant
pressure $p$ and density $\mu$.}\\[2mm]
{\it Proof.}
The equations  $\Psi_0=\Psi_2=0$, with (\ref{psi0}) and (\ref{psi2})
yield $E_2=H_3=-E_1$ and $E_3=-H_2=H_1$. We next arrange 
$\omega_2+\sigma_2= 0$ by using the transformation (\ref{transf}).
 We obtain $H_1$ and $E_1$ from
the algebraic equations (\ref{rr6}) and (\ref{rr8}), respectively.
The equation of state $\mu+p=0$ follows at once from (\ref{rr11}).
From the Bianchi identities (\ref{bi1}) and (\ref{bi2}) the statement
of the theorem follows.  \hfill
$\Box$\\[2mm]

\subsection{Rigidly rotating electric space-times}

In this subsection we consider rigidly rotating fluid spacetimes with
purely electric Weyl tensor. We then have $\sigma_1=\sigma_2=0$ and
$H_A=0$. These spacetimes have the property stated in the
following \\[2mm]
{\bf Theorem 4.}
{\it A rigidly rotating purely electric perfect fluid space-time is of
 Petrov type D. The four-velocity lies in the plane spanned by the
 principal null directions of the Weyl tensor.
}\\[2mm]
{\it Proof.}
We set $\omega_1 = 0$ by using the transformation (\ref{transf}).
From the Bianchi equations (\ref{b5}), (\ref{b7}) and (\ref{b8})
we get, respectively
\begin{align}
&E_3=0 \\
&E_2=-2E_1 \\
&\mu=3E_2+p \ .
\end{align}
The combination of Bianchi equations (\ref{b1})--(\ref{b4}) yields
\begin{align}
&\alpha_2=0 \\
&\beta_2=-\alpha_1 \ .
\end{align}
From the Ricci equations (\ref{rr16}) and (\ref{rr15}) we have,
respectively
\begin{align}
&\alpha_1=a_2 \\
&E_1=\frac12(a_1\beta_1-a_2^2-\omega_2^2)+\frac16 p \ .
\end{align}
From conditions (\ref{typeD1}) and (\ref{typeD2}) it follows immediately that
these fields are of type D. The double principal null directions are
${\bf l} + {\bf n} + i({\bf m - \bar m})$ and
${\bf l} + {\bf n} - i({\bf m - \bar m})$. Hence
the four-velocity $\bfe_0$ lies
in the plane of the principal null directions.\hfill
$\Box$\\[2mm]
The condition that the
rigidly rotating perfect fluid space-time is Petrov type D and that the
four-velocity is  a linear combination of the principal directions has been
thoroughly investigated by Senovilla\cite{Senovilla}, who gave all the exact
solutions in this class.

\section{Acknowledgment}
This work has been supported by OTKA fund T022533,
the Royal Swedish Academy of Sciences, and the Hungarian
Academy of Sciences. Thanks are due to C.
McIntosh for calling our attention to magnetic fluids.


\begin{thebibliography}{99}
  \bibitem{DeFelice-Clarke}
    De Felice, F.\ and C.\ J.\ S.\ Clarke (1990) \textit{Relativity on
    curved manifolds}, pp.\ 254--255 (Cambridge: Cambridge University
    Press).

  \bibitem{Arianrhod}
    Arianrhod {\em et al.}\ (1994) Magnetic curvatures,
    \textit{Class.\ Quantum Grav.}\ \textbf{11}, 2331--2335.

  \bibitem{Misra}
    Misra {\em et al.}\ (1968) A solution of the field
    equations representing a gravitational field of a magnetic type,
    \textit{Tensor}\ \textbf{19}, 203--205.

  \bibitem{McIntosh}
    McIntosh {\em et al.}\ (1994) Electric and magnetic Weyl tensors:
    Classification and analysis,
    \textit{Class.\ Quantum Grav.}\textbf{11}, 1555.

  \bibitem{Bradley-Karlhede}
    Bradley, M.\ and A.\ Karlhede (1991) On the curvature description
    of gravitational fields, \textit{Class.\ Quantum Grav.}\
    \textbf{7} 449--463.

  \bibitem{Bradley-Marklund}
    Bradley, M.\ and M.\ Marklund (1996) Finding solutions to
     Einstein's equations in terms of invariant objects,
    \textit{Class.\ Quantum Grav.}\ \textbf{13} 3021--3037.

  \bibitem{Collinson}
    Collinson, C.\ D.\ (1976) The Uniqueness of the Schwarzschild
    Interior Metric, \textit{Gen.\ Rel.\ Grav.}\ \textbf{7} 419--422.

  \bibitem{Newman}
    Luk\'acs, B. {\it et al.} (1983) A NUT-like solution with fluid
    matter, \textit{Gen.\ Rel.\ Grav.}\ \textbf{15} 567--571.


  \bibitem{Senovilla}
    Senovilla, J.M.M. (1987) On Petrov type-D stationary axisymmetric
    rigidly rotating perfect fluid metrics, \textit{Class.\ Quantum
    Grav.}\ \textbf{4} L115--L119.
\end{thebibliography}
\end{document}